\documentclass{gji}

\usepackage{timet,color}
\usepackage{natbib} 
\usepackage{graphicx}
\usepackage{amsmath,amssymb}
\usepackage[urlcolor=blue,citecolor=black,linkcolor=black]{hyperref}

\providecommand{\newblock}{}

\title[Ionospheric disturbances of the 2025 Myanmar earthquake] 
  {Cross-sphere Coupling and Source Inversion of Ionospheric Disturbances Associated with the 2025 Myanmar Strike-slip Earthquake from BeiDou GEO and Multi-GNSS Observations}

\author[J. Chen et al.]
  {Jianghe Chen$^{1}$, Pan Xiong$^{1}$\thanks{Corresponding author: xiongpan@ief.ac.cn}, Qingshan Ruan$^{2}$, Xiaoran Zhang$^{1}$, Yuqi Lin$^{1}$,
  \and Xuemin Zhang$^{1}$, Ting Zhang$^{3}$, Kaixin Wang$^{4}$, and Xuhui Shen$^{5}$\thanks{Corresponding author: shenxuhui@nssc.ac.cn}\\
   $^{1}$Institute of Earthquake Forecasting, China Earthquake Administration, Beijing 100036, China\\
   $^{2}$Hanzhong Vocational and Technical College, Hanzhong 723000, China\\
   $^{3}$Innovation Academy for Precision Measurement Science and Technology, Chinese Academy of Sciences, Wuhan 430071, China\\
   $^{4}$School of Civil Engineering and Geomatics, Shandong University of Technology, Zibo 255000, China\\
   $^{5}$National Space Science Center, Chinese Academy of Sciences, Beijing, China
  }
\date{Received 2025 XXX XX; in original form 2025 XXX XX}
\pagerange{\pageref{firstpage}--\pageref{lastpage}}
\volume{XXX}
\pubyear{{\the\year{}}}

\begin{document}

\label{firstpage}
\maketitle


\begin{summary}
Focusing on the M7.9 earthquake in Myanmar in 2025, this study comprehensively utilizes data from BeiDou geostationary satellites of the Chinese Continental Crustal Movement Observation Network and multi-system Global Navigation Satellite Systems (GNSS). The spatiotemporal evolution characteristics and physical mechanisms of pre-seismic ionospheric anomalies and co-seismic ionospheric disturbances were systematically analyzed. By employing the moving interquartile range method combined with solar-terrestrial environmental parameters, a negative Total Electron Content (TEC) anomaly associated with the seismogenic region was identified three days before the earthquake. The equatorial conjugate structure of this TEC anomaly revealed a multi-path coupling effect between the lithosphere, atmosphere, and ionosphere. The extraction of Coherent Ionospheric Disturbance (CID) signals based on wavelet transform and band-pass filtering indicated that the co-seismic ionospheric disturbances were dominated by acoustic-gravity waves in the 2-8 mHz frequency band, propagating at a speed of approximately 1.2 km/s, and exhibiting an asymmetric pattern in the southeast direction. A spatial density-weighted method for locating the source of ionospheric disturbances was proposed, elucidating the joint control mechanism of fault strike-slip motion, geomagnetic field modulation, and equatorial electrojet on the disturbance energy. The results confirm that the high spatiotemporal resolution of BeiDou GEO satellites and multi-system GNSS significantly enhances the capability to capture weak ionospheric anomaly signals associated with earthquakes. These results provide additional observational constraints on space-based Lithosphere-Atmosphere-Ionosphere Coupling (LAIC) processes and may contribute to the development of ionosphere-based earthquake monitoring techniques.
\end{summary}




\section{Introduction}
The Lithosphere-Atmosphere-Ionosphere Coupling (LAIC) mechanism is a crucial direction in the study of earthquake precursors. The core of this mechanism lies in the physical and chemical anomalies at the Earth's surface, triggered by stress accumulation in the seismogenic region, which may influence the ionospheric plasma distribution through various pathways such as electromagnetic radiation, acoustic-gravity waves, and aerosol diffusion. As a direct manifestation of intense crustal activity, major earthquakes provide a natural experimental setting for validating LAIC mechanisms \citep{Afraimovich2010,Astafyeva2011,Heki2024a,Heki2015,Jia2024,Liu2024,Shah2020,Shi2021,Song2022,Tang2024,Yao2012}. However, traditional GPS/GNSS-based Total Electron Content (TEC) monitoring relies on signals from medium-Earth-orbit (MEO) satellites, whose ionospheric pierce points (IPPs) migrate rapidly over the ground. This moving-IPP geometry introduces strong spatiotemporal aliasing, making it difficult to distinguish the temporal evolution and spatial gradients of ionospheric anomalies. This challenge is exacerbated during periods of high solar activity, when the background noise in the ionosphere increases, further reducing the signal-to-noise ratio of anomaly signals \citep{Feng2023,Matamba2023,Zhu2022}.

At low latitudes, the daytime eastward equatorial electrojet (EEJ) and the E$\times$B fountain effect generate the Equatorial Ionization Anomaly (EIA), with enhanced F-region electron density on both sides of the magnetic equator. In typical conditions, the crests of the EIA are located at magnetic latitudes of order 10--20$^\circ$ on either side of the magnetic equator and reach their maximum intensities during the afternoon to evening sector (approximately 14--20 local time) \citep[e.g.][]{Abdu2005}. This background structure and the associated plasma drifts strongly modulate the propagation and amplitude of seismo- and tsunami-induced ionospheric disturbances, and thus need to be taken into account when interpreting case studies of earthquake-related ionospheric signals.

In recent years, the networking operation of Chinese BeiDou geostationary satellites has provided a new approach to addressing these issues. The fixed pierce point observation characteristics of BeiDou GEO satellites enable second-level continuous monitoring of TEC in specific regions, effectively eliminating spatial sampling errors. This provides a unique advantage for extracting weak pre-seismic anomaly signals and capturing the dynamic propagation processes of co-seismic ionospheric disturbances (CID) \citep{Chen2024,Jia2024,Tang2022,Li2023,Chen2022}. The M7.9 Myanmar earthquake on March 28, 2025, occurred along the right-lateral Sagaing Fault system in central Myanmar, which accommodates a large portion of the relative motion between the Burma microplate and the Sunda Plate at the eastern boundary of the India--Eurasia collision zone. This region exhibits strong crustal deformation rates and frequent large earthquakes, making it an ideal natural laboratory for studying LAIC processes associated with plate-boundary strike-slip events.

This study integrates data from BeiDou GEO satellites and multi-system GNSS networks, combining wavelet power spectrum analysis and weighted grid inversion methods. It reveals the key controlling factors of energy transfer in the LAIC mechanism of strike-slip earthquakes from four dimensions: spatiotemporal correlation of pre-seismic ionospheric anomalies, spectral characteristics of disturbances, azimuthal propagation asymmetry, and source location error constraints. For the pre-seismic phase, continuous TEC time series data from BeiDou GEO satellites were used, employing the moving interquartile range method combined with solar activity parameters to remove background noise. A significant negative TEC anomaly was identified three days before the earthquake, centered within the Dobrovolsky seismogenic radius ($R \approx 1740$ km). The anomaly exhibited equatorial conjugate structure evolution, providing observational evidence for the cross-sphere coupling between fault-controlled crustal processes and ionospheric disturbances. By jointly analyzing the preseismic anomalies and co-seismic CIDs within a unified framework, we propose a consistent case-based picture of LAIC energy transfer for this large strike-slip earthquake.

In this study we first describe the earthquake and GNSS data (Section~2), then present the preseismic anomaly and coseismic CID analyses (Section~3), followed by a discussion of the underlying LAIC processes (Section~4) and conclusions (Section~5).

\section{Data and Methods}
\subsection{Earthquake Case and GNSS Data}
According to the China Earthquake Networks Center, a magnitude 7.9 earthquake occurred in Myanmar (21.85°N, 95.95°E) at 14:20 on March 28, 2025, with a focal depth of 30 km. This M7.9 earthquake occurred along the right-lateral Sagaing Fault system in central Myanmar, approximately 294 km from the Chinese border. Since the Cenozoic era, the collision between the Indian and Eurasian plates has led to large-scale shortening of the lithosphere, forming the Himalayan orogenic belt and the adjacent Burma microplate. The Myanmar Arc region accommodates part of the relative motion between the Burma microplate and the Sunda Plate and is characterized by strong crustal deformation and high seismic activity. It hosts a series of north--south trending tectonic units and faults, including the Sagaing Fault, which ruptured during this earthquake. The Sagaing Fault extends from northern Myanmar near the eastern Himalayan front southward along the Irrawaddy River to Sagaing, then continues along the Sittang River to Bago. After entering the sea, it extends along the western side of the Mergui Archipelago and becomes unclear near Phuket, Thailand. The total length of the Sagaing Fault is about 1200 km, and it is a right-lateral strike-slip fault with an average annual slip rate of 18--23 mm. This region is one of the most seismically active areas along the Himalayan seismic belt. Since 1900, there have been 10 earthquakes of magnitude 7 or greater within a 300 km radius of the epicenter, with the largest being an M8.0 earthquake on May 23, 1912, in Myanmar, which caused significant landslides and other secondary hazards, resulting in substantial casualties. The regional stress field, dominated by the northeastward compression from the Indian plate, results in a predominantly right-lateral strike-slip regime along the Sagaing Fault, consistent with the focal mechanism of the studied earthquake.

\begin{figure*}
\centering
\includegraphics[width=\textwidth]{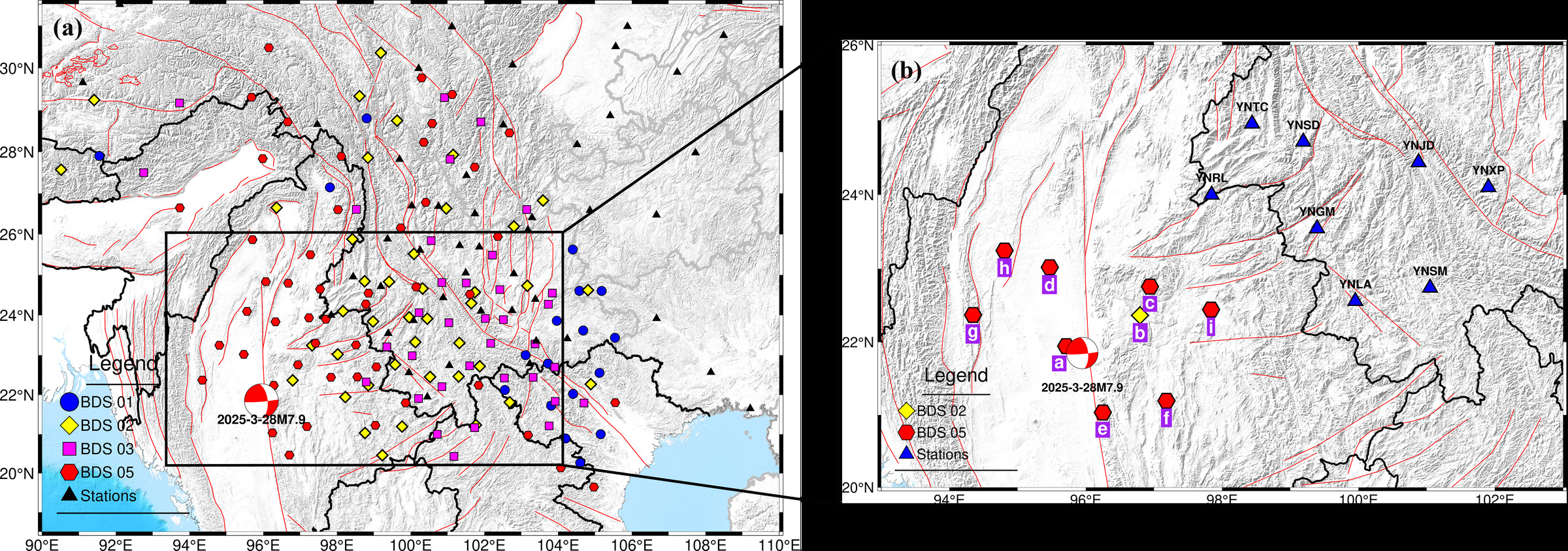}
\caption{Distribution of BeiDou geostationary satellite ionospheric pierce points (IPPs) in the study area. The labels (a)--(i) mark the nine representative IPPs used in the preseismic time-series analysis in Figure~\ref{fig:pre_seismic_tec}; the remaining symbols show the full set of GEO IPPs in the study region.}
\label{fig:ipp_distribution}
\end{figure*}

The data for this study were sourced from the Crustal Movement Observation Network of China (CMONOC). The raw GNSS observation files were processed to generate slant TEC (STEC) values with a 30-second sampling interval. We applied a cycle-slip detection and correction algorithm and set a satellite elevation mask angle of 15° to mitigate multipath effects and atmospheric noise at low elevations. For the preseismic anomaly analysis, we selected TEC data from nine representative ionospheric pierce points (IPPs; labelled a--i in Figure~\ref{fig:ipp_distribution}) corresponding to eight ground observation stations near the projected fault trace. Other IPPs shown in Figure~\ref{fig:ipp_distribution} illustrate the overall spatial coverage of the BeiDou GEO observations but were not used in the sliding-window anomaly detection. The Geostationary Earth Orbit (GEO) satellites of the BeiDou Navigation Satellite System (BDS) exhibit significant advantages in TEC monitoring, primarily due to their unique orbital characteristics:
\begin{itemize}
\item Fixed ionospheric pierce points (IPPs): BeiDou GEO satellites are geostationary relative to the Earth, meaning their ionospheric pierce points remain fixed. This allows for long-term continuous observations of the same region, effectively eliminating spatial sampling errors in TEC data.
\item Long-term 24/7 continuous observations: BeiDou GEO satellites provide continuous signal coverage over specific regions (e.g., seismic zones) on diurnal to multi-year timescales, enabling the generation of high-temporal-resolution TEC time series and providing a stable data foundation for high-precision ionospheric modeling.
\item High smoothness and signal-to-noise ratio (SNR): The fixed-point observation characteristics of BeiDou GEO satellites avoid ionospheric spatial gradient errors caused by satellite motion, facilitating the extraction of weak anomaly signals.
\end{itemize}

\subsection{Pre-Seismic Anomaly Detection}
The moving interquartile range method was used for pre-seismic ionospheric anomaly detection, with a sliding window length of 27 days (corresponding to the solar rotation period) to eliminate the influence of normal diurnal ionospheric periodic variations. For each IPP, we computed the running median and interquartile range (IQR) within the 27-day window and defined a prediction interval as median $\pm$ 1.5~$\times$~IQR. TEC values falling above (below) this interval were flagged as ``above-threshold'' (``below-threshold'') anomalies in the subsequent analysis. Given that the current period is the peak year of the 25th solar activity cycle, the ionospheric background noise is significantly enhanced. This study combined data from the ``Space Environment Weekly Report'' issued by the Chinese Academy of Sciences' Space Environment Research and Forecast Center to simultaneously analyze solar wind speed, interplanetary magnetic field strength, and geomagnetic activity indices (Kp, Dst). This approach allows for the effective identification and removal of ionospheric disturbances caused by solar activity.

\subsection{Co-Seismic Disturbance Analysis}
\begin{figure*}
\centering
\includegraphics[width=\textwidth]{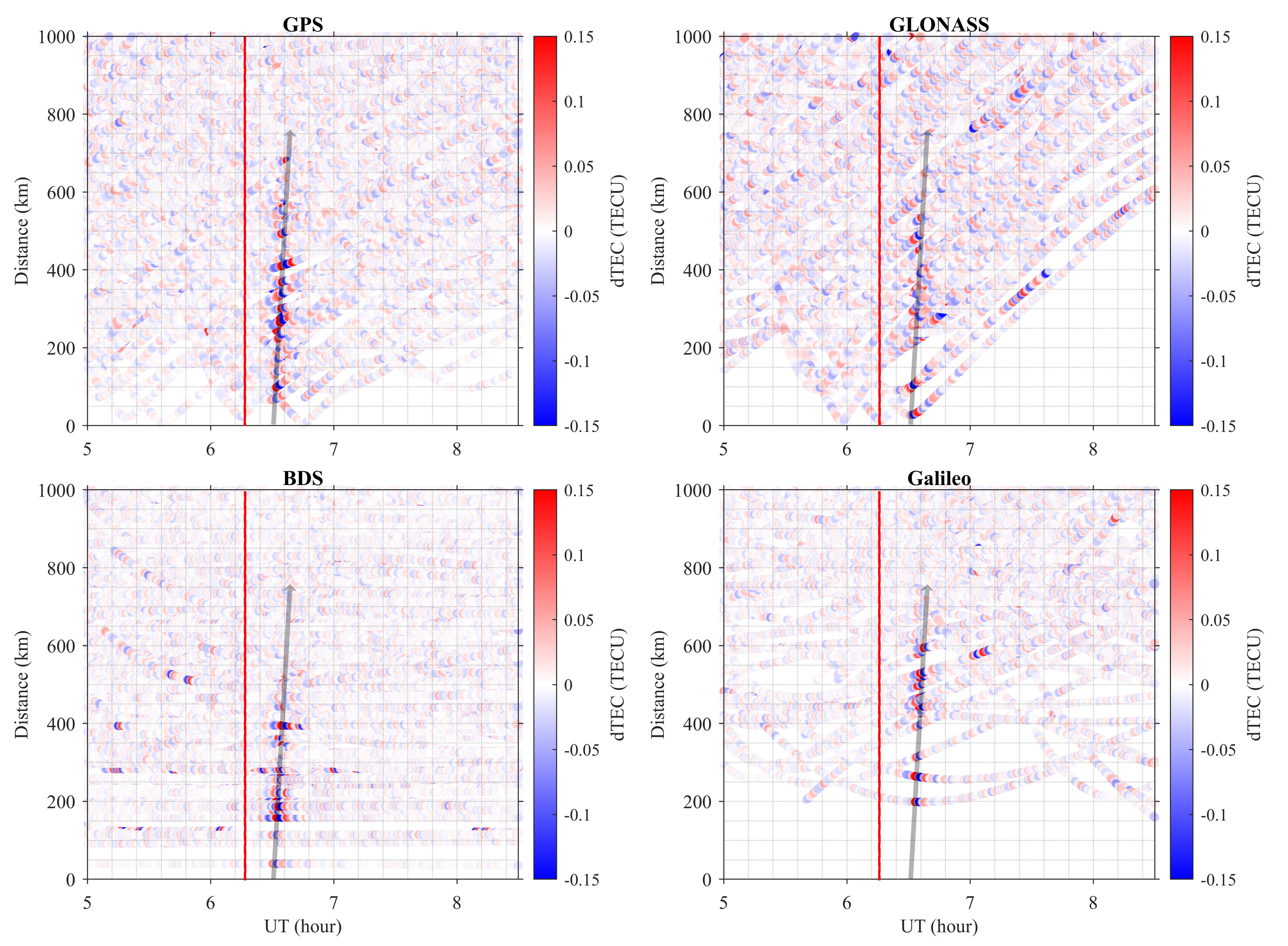}
\caption{Time--distance plots showing co-seismic ionospheric disturbance propagation velocities derived from multi-system GNSS TEC data. The black arrows indicate the apparent propagation directions and velocities and have been drawn with reduced thickness to avoid obscuring the TEC disturbance signals.}
\label{fig:cid_multisystem_velocity}
\end{figure*}

\begin{figure*}
\centering
\includegraphics[width=\textwidth]{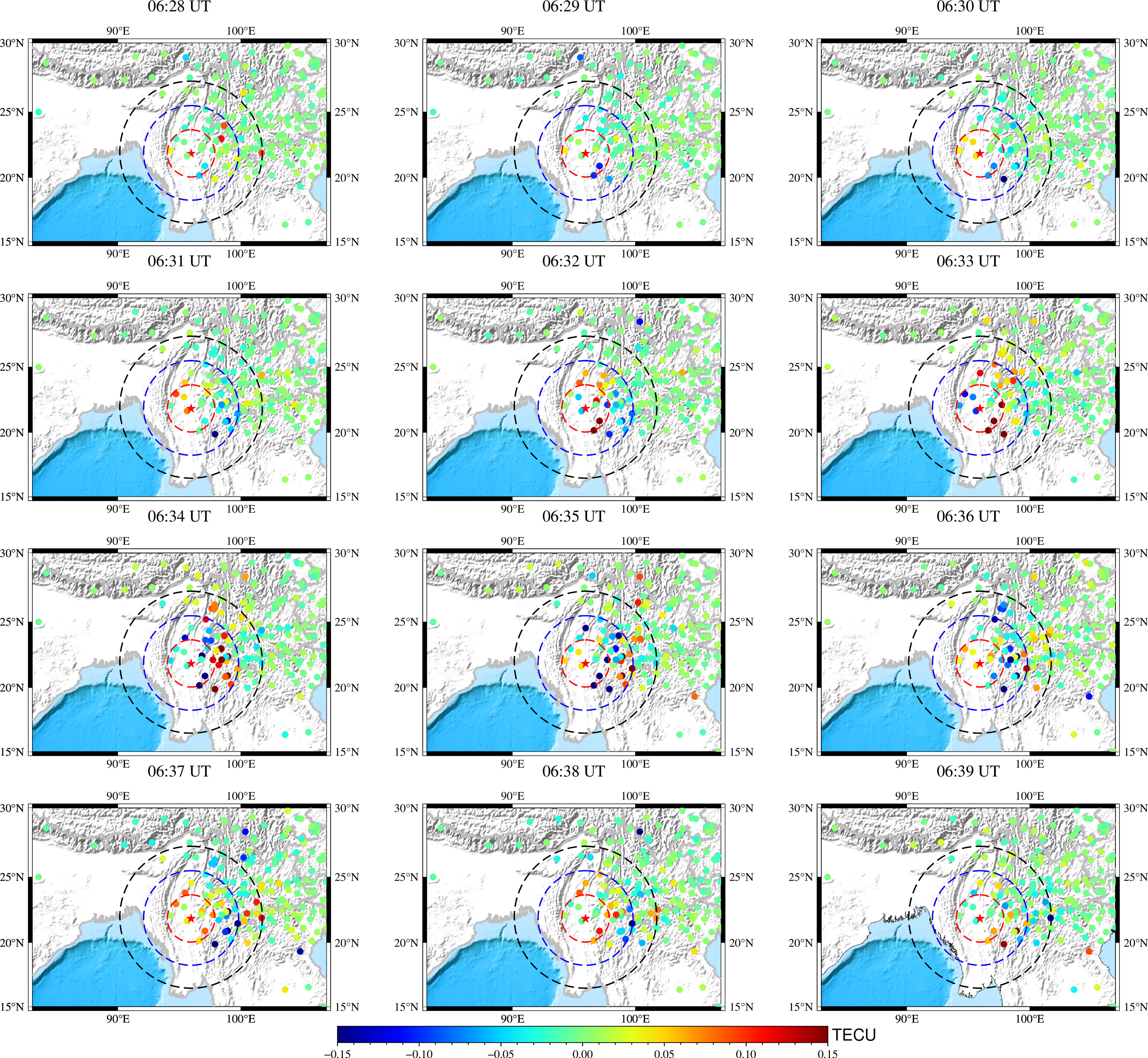}
\caption{Azimuthal distribution of co-seismic ionospheric disturbance points observed in TEC derived from BeiDou GEO satellite solutions. The red star marks the epicenter. The dashed circles indicate areas within 200 km (red), 400 km (blue), and 600 km (black) of the earthquake center.}
\label{fig:cid_azimuth_geo}
\end{figure*}

\begin{figure*}
\centering
\includegraphics[width=\textwidth]{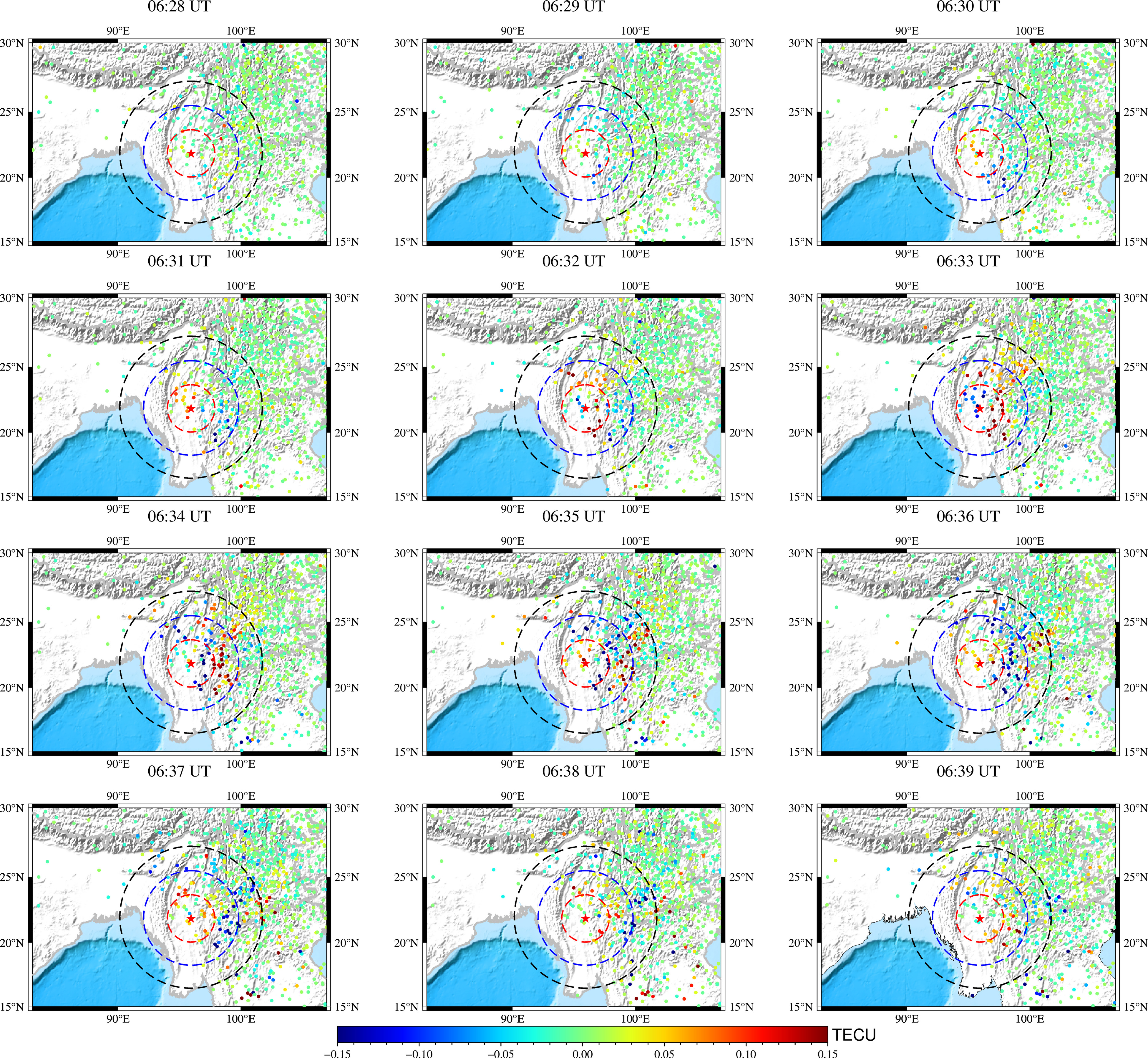}
\caption{Azimuthal distribution of co-seismic ionospheric disturbance points observed in TEC derived from multi-system GNSS solutions. Symbols are the same as in Figure~\ref{fig:cid_azimuth_geo}.}
\label{fig:cid_azimuth_multisystem}
\end{figure*}

\begin{figure*}
\centering
\includegraphics[width=\textwidth]{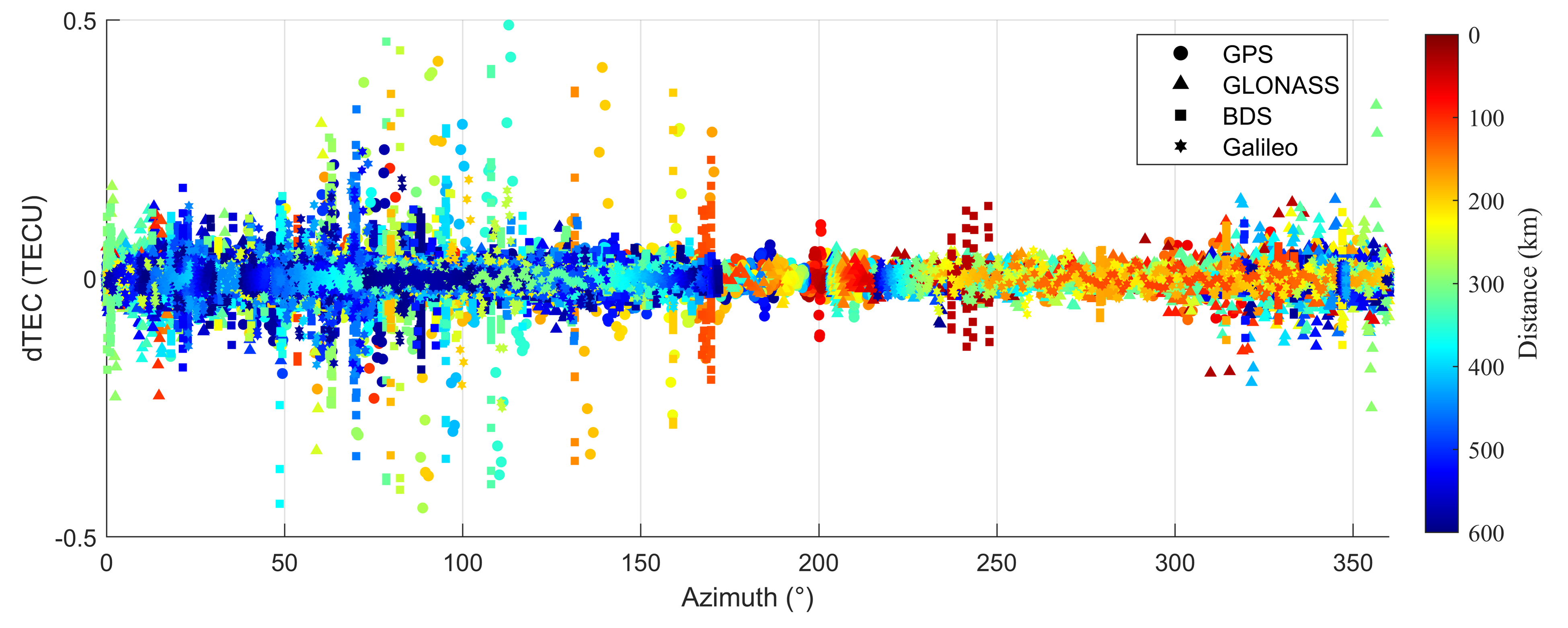}
\caption{Additional observations of co-seismic ionospheric disturbances, showing TEC variation as a function of time and azimuth.}
\label{fig:cid_azimuth_extra}
\end{figure*}

To study the characteristics of co-seismic ionospheric disturbances (CIDs), we first removed the slowly varying background from the TEC time series to obtain differential TEC (DTEC). We then employed wavelet transform to analyze the detrended (but not band-pass filtered) DTEC series in the time--frequency domain \citep{Ruan2023}. The wavelet spectra reveal a significant energy concentration in the 2--8 mHz frequency band. Guided by this result, we applied a Butterworth band-pass filter with cutoff frequencies of 2 mHz and 8 mHz to the original TEC observations to effectively extract the ionospheric disturbance signals related to the earthquake. This processing method retains the essential characteristics of the CIDs while eliminating high-frequency noise and low-frequency background variations. The 2--8 mHz band-pass filtered TEC is used for the time--distance representation in Figure~\ref{fig:cid_bds_geo}(c) and for the subsequent time--distance and azimuthal analyses (Figures~\ref{fig:cid_multisystem_velocity}--\ref{fig:cid_azimuth_extra}). For azimuthal analysis, we also filtered TEC data from multi-system GNSS (GPS, BDS, Galileo, GLONASS) observations from CMONOC using the same 2--8 mHz band-pass filter.
We verified that the 2--8 mHz band-pass filtering does not introduce significant phase shifts within the passband, so the inferred propagation velocities and onset times are not biased by the filter design.

\subsection{CID Source Inversion Method}
This study adopted a grid search inversion method with spatial density weighting to determine the source location of the earthquake-induced ionospheric disturbances. The specific procedures are as follows. First, the original data of ionospheric Total Electron Content disturbances (dTEC) were screened, and only the valid data with an observation duration of $\leq 600$ s were retained. The data were grouped according to satellite--receiver combinations, and the disturbance event with the maximum absolute value of dTEC in each group was extracted as a characteristic point. For each characteristic point $i$, we recorded the geographical coordinates of the corresponding ionospheric pierce point (IPP) and the observed arrival time of the first CID peak $t_{\mathrm{obs},i}$. In total, $N$ valid IPP samples were obtained.

The ionosphere was approximated by a thin shell at a fixed height $h_i = 350$ km, and the vertical group velocity of the CID-related acoustic--gravity waves was taken as $v_z = 0.5$ km s$^{-1}$, following previous studies \citep[e.g.][]{He2017,Ruan2023}. The horizontal propagation velocity $v_h$ was treated as an unknown parameter to be determined. A search grid of $0.1^\circ \times 0.05^\circ$ was constructed in the potential source area (latitude $18^\circ$--$28^\circ$, longitude $90^\circ$--$102^\circ$). The grid resolution of $0.1^\circ \times 0.05^\circ$ was chosen as a balance between computational efficiency and localization precision, corresponding to a spatial resolution of approximately 11 km $\times$ 5.5 km in the study region.

For each trial source location $\mathbf{x}$ on the grid and each candidate horizontal propagation speed $v_h$, we computed the great-circle horizontal distance $d_i(\mathbf{x})$ between $\mathbf{x}$ and IPP $i$ (in km, using the Haversine formula). The corresponding travel-time residual for IPP $i$ was defined as
\begin{equation}
\Delta t_i(v_h,\mathbf{x}) = t_{\mathrm{obs},i} - \left( \frac{h_i}{v_z} + \frac{d_i(\mathbf{x})}{v_h} \right),
\end{equation}
which follows the standard travel-time relation used in CID source inversion \citep[e.g.][]{He2017,Ruan2023}.

To account for both temporal consistency and spatial sampling, we constructed a composite weight system. A time weight was defined as
\begin{equation}
w_{t,i} = \exp\left(-\frac{|\Delta t_i|}{1000}\right),
\end{equation}
where the characteristic time scale of 1000 s gives higher influence to observations whose arrival times are more consistent with the trial source while smoothly down-weighting outliers. A spatial density weight was defined as
\begin{equation}
w_{s,i} = \frac{1}{\rho_i + 1},
\end{equation}
where $\rho_i$ is the local density of IPPs (the number of IPPs within the grid cell containing IPP $i$); the addition of 1 avoids singularities where the density is zero. The weights were normalized to ensure comparability under different parameters,
\begin{equation}
w_{s,i}^{\mathrm{norm}} = \frac{w_{s,i}}{\max(w_s)}.
\end{equation}

For each trial $(v_h,\mathbf{x})$, the total weight for IPP $i$ is $w_i = w_{t,i}\,w_{s,i}^{\mathrm{norm}}$, and the weighted standard deviation of the travel-time residuals is calculated as
\begin{equation}
\mathrm{STD}(v_h,\mathbf{x}) = \sqrt{\frac{\sum_{i=1}^{N} w_i \left[\Delta t_i(v_h,\mathbf{x}) - \overline{\Delta t}(v_h,\mathbf{x})\right]^2}{\sum_{i=1}^{N} w_i}},
\end{equation}
where $\overline{\Delta t}(v_h,\mathbf{x})$ is the weighted mean residual. STD (in seconds) measures the internal consistency between the modeled and observed arrival times: smaller values indicate better agreement. After traversing all grid points, the position with the smallest STD is selected as the optimal apparent source location corresponding to the current $v_h$. The final $v_h$ value was selected based on the combination that resulted in the minimum localization error.

\section{Results}
\subsection{Pre-Seismic Ionospheric Anomalies}
\begin{figure*}
\centering
\includegraphics[width=0.55\textwidth]{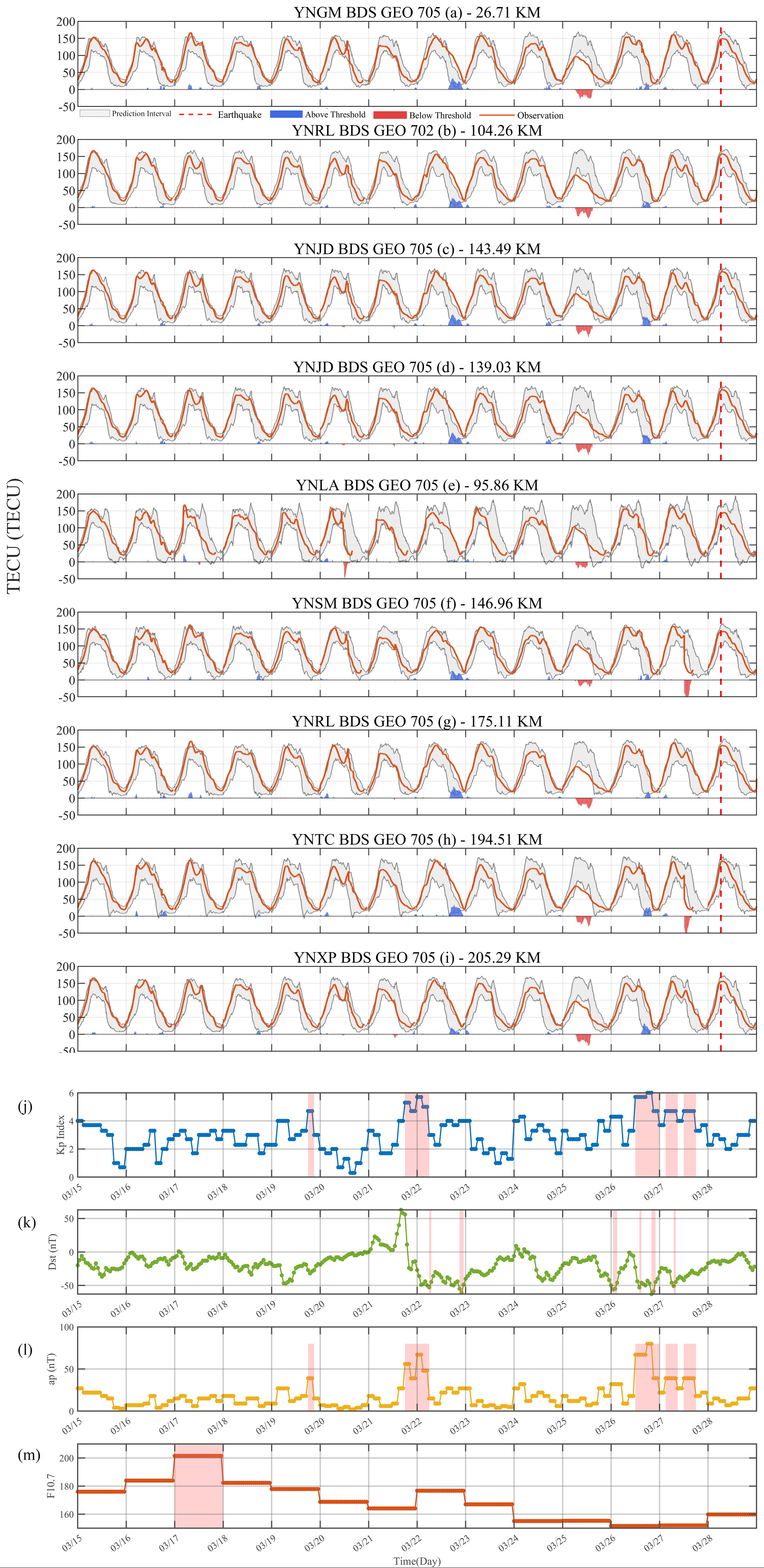}
\caption{Ionospheric TEC anomalies detected by the BeiDou geostationary satellites before the earthquake. Panels (a)--(i) show detrended TEC time series at the nine representative IPPs. The red line in each panel denotes the 27-day running median, and the shaded band shows the prediction interval (median $\pm$ 1.5~$\times$~IQR). Blue (red) markers indicate data points below (above) this interval. The number followed by ``km'' in each panel title denotes the horizontal distance between the corresponding IPP and the earthquake epicenter. Panels (j)--(m) display relevant space environment indices, and the light pink shaded areas indicate periods of disturbed solar or geomagnetic activity. Panels (a)--(i) correspond to the IPPs labelled (a)--(i) in Figure~\ref{fig:ipp_distribution}.}
\label{fig:pre_seismic_tec}
\end{figure*}

\begin{figure*}
\centering
\includegraphics[width=\textwidth]{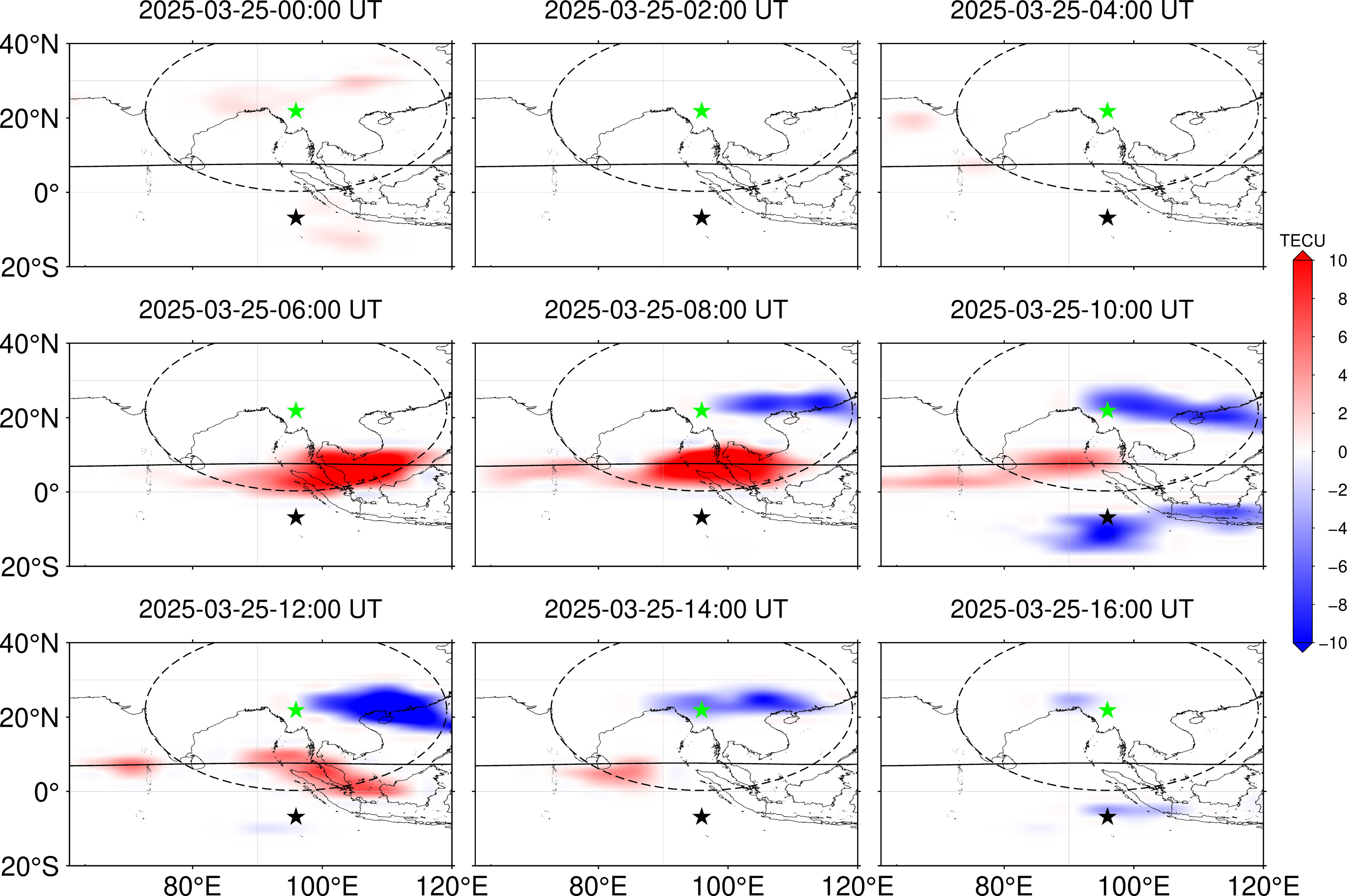}
\caption{Distribution of JPL Global Ionosphere Map (GIM) TEC anomalies on March 25. The green star marks the earthquake epicenter, and the black star indicates the magnetic equatorial conjugate point of the epicenter. The solid black line between them represents the magnetic equator.}
\label{fig:jpl_tec_anomaly}
\end{figure*}

On March 25 (three days before the earthquake), under stable geomagnetic conditions, a significant negative TEC anomaly, reaching a minimum of -2.5 TECU below the median, was observed simultaneously at all nine IPPs at 12:00 UT (Figure~\ref{fig:pre_seismic_tec}). This phenomenon contradicts the typical ionospheric enhancement characteristics during the peak of solar activity. To investigate the spatial distribution characteristics of the anomaly, we utilized the Global Ionosphere Map (GIM) and applied the same statistical method. The results show that the anomalous region was mainly concentrated in the vicinity of the Myanmar epicenter, covering an approximate area of 500,000 km² (Figure~\ref{fig:jpl_tec_anomaly}).

The evolution of the ionospheric anomaly on March 25 exhibited significant spatiotemporal correlation lasting approximately 10 hours. At 06:00 UT, a positive TEC anomaly was detected in the southeastern quadrant of the seismogenic region. The center of this anomalous region was approximately 850 km from the epicenter, within the theoretical Dobrovolsky seismogenic radius ($R = 10^{(0.43M)} \approx 1740$ km for M7.9). By 08:00 UT, the anomaly system showed significant spatial differentiation: a new positive anomaly core formed south of the epicenter, while an extended negative anomaly appeared in the northeastern quadrant. At 10:00 UT, a cross-equatorial conjugate phenomenon was observed: the conjugate point (20°S, 100-120°E) of the original negative anomaly in the northeastern quadrant (20°N, 100-120°E) also showed a negative anomaly, with a positive anomaly band appearing in the equatorial trough between them. After 12:00 UT, the anomaly system entered a decay phase, and by 16:00 UT, it had completely dissipated, completing the "positive-negative alternation-conjugate response-decay" temporal sequence.

We also examined the anomaly flags within the 27-day sliding window preceding 25 March and did not find negative TEC events with comparable spatial extent and coherence. This suggests that the March 25 anomaly was unusual relative to the short-term background, although a full statistical assessment over longer time periods and multiple events is beyond the scope of this study.

\subsection{Characteristics of Co-Seismic Ionospheric Disturbances}
\begin{figure*}
\centering
\includegraphics[width=\textwidth]{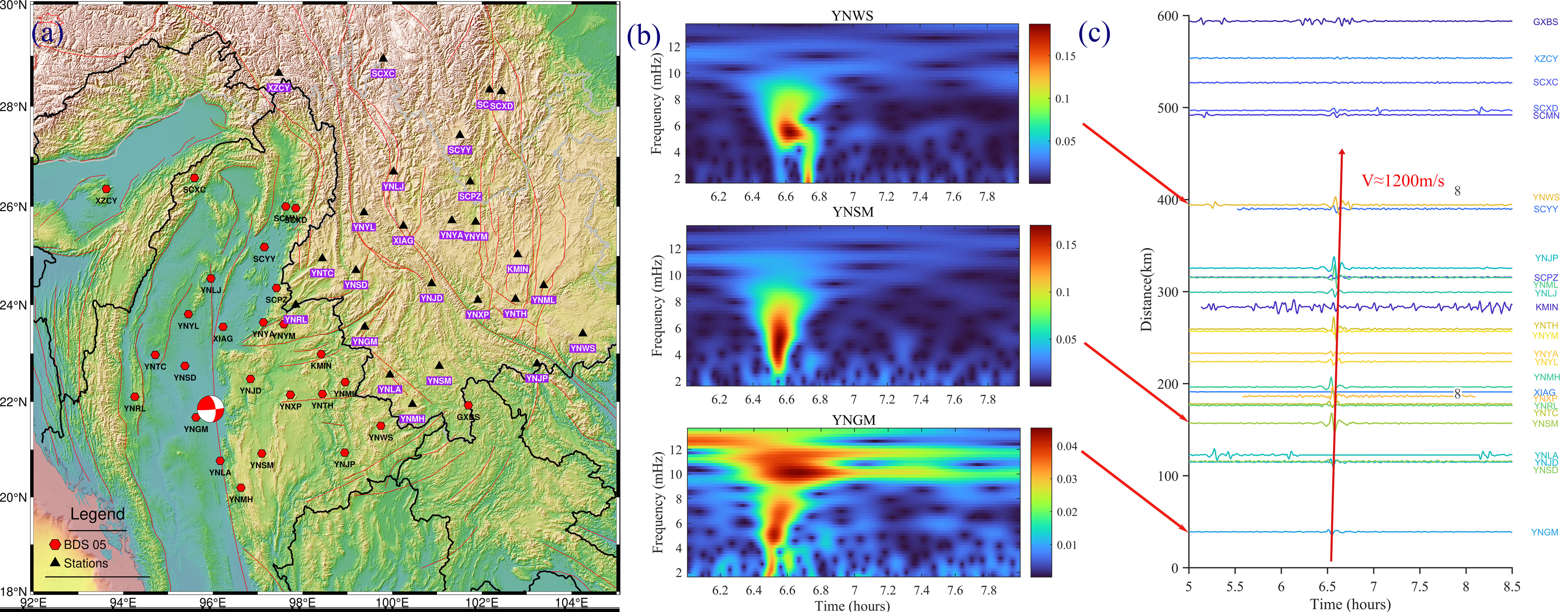}
\caption{(a) Location of the earthquake, with black triangles representing CMONOC sites and red hexagons representing the IPPs of the corresponding BeiDou-5 GEO satellite. (b) Detrended TEC time series (without band-pass filtering) from the BeiDou-5 GEO satellite for the YNWS, YNSM, and YNGM stations together with their wavelet power spectra. The 2--8 mHz band (indicated by dashed lines) is used to design the Butterworth band-pass filter applied in panel (c) and subsequent figures. (c) Time--distance plot of co-seismic ionospheric disturbances detected by the BeiDou-5 GEO satellite, based on 2--8 mHz band-pass filtered TEC.}
\label{fig:cid_bds_geo}
\end{figure*}

The wavelet power spectrum analysis of DTEC time series for three Beidou GEO satellite IPPs near the epicenter (YNGM, YNSM, YNWS) revealed a significant energy concentration in the 2-8 mHz frequency band for all three IPPs (Figure~\ref{fig:cid_bds_geo}b). As shown in Figure~\ref{fig:cid_bds_geo}(b), the wavelet power spectra for all three stations show a distinct enhancement in the 2–8 mHz band beginning around 06:30 UTC, approximately 9–10 minutes after the mainshock. The signal at the closest station, YNGM, exhibits the highest power and persists for over 30 minutes, while the signal at the more distant YNWS station is weaker and has a shorter duration. The YNGM station, closest to the epicenter, also showed a high-energy region around 10 mHz.

The first CID signal was detected approximately 9 minutes after the main shock. Time-distance plots derived from both Beidou GEO data (Figure~\ref{fig:cid_bds_geo}c) and multi-system GNSS data (Figure~\ref{fig:cid_multisystem_velocity}) show that the CID propagated at a stable speed of approximately 1.2 km/s (or 1200 m/s). The disturbance remained significant up to 400 km from the epicenter. Comparative analysis of the four GNSS systems indicates that GPS, BDS, and Galileo clearly captured the disturbance event with a peak CID amplitude of approximately 0.4-0.5 TECU, while the GLONASS signal was less apparent, showing a peak of only 0.1 TECU.

\subsection{Azimuthal Asymmetry of CIDs}
Figures~\ref{fig:cid_azimuth_geo}, \ref{fig:cid_azimuth_multisystem}, and \ref{fig:cid_azimuth_extra} show the filtered TEC variation with azimuth angle. The analysis reveals a very significant anomaly in the southeast direction (azimuth angles 90°-150°). Between 06:29 and 06:31, a strong negative anomaly of up to -0.5 TECU appeared first in the southeast direction, while a weaker positive anomaly of approximately +0.15 TECU was observed in the northwest, indicating that the CIDs generated by this earthquake were anisotropic.

\subsection{Source Inversion of CIDs}
\begin{figure*}
\centering
\includegraphics[width=0.7\textwidth]{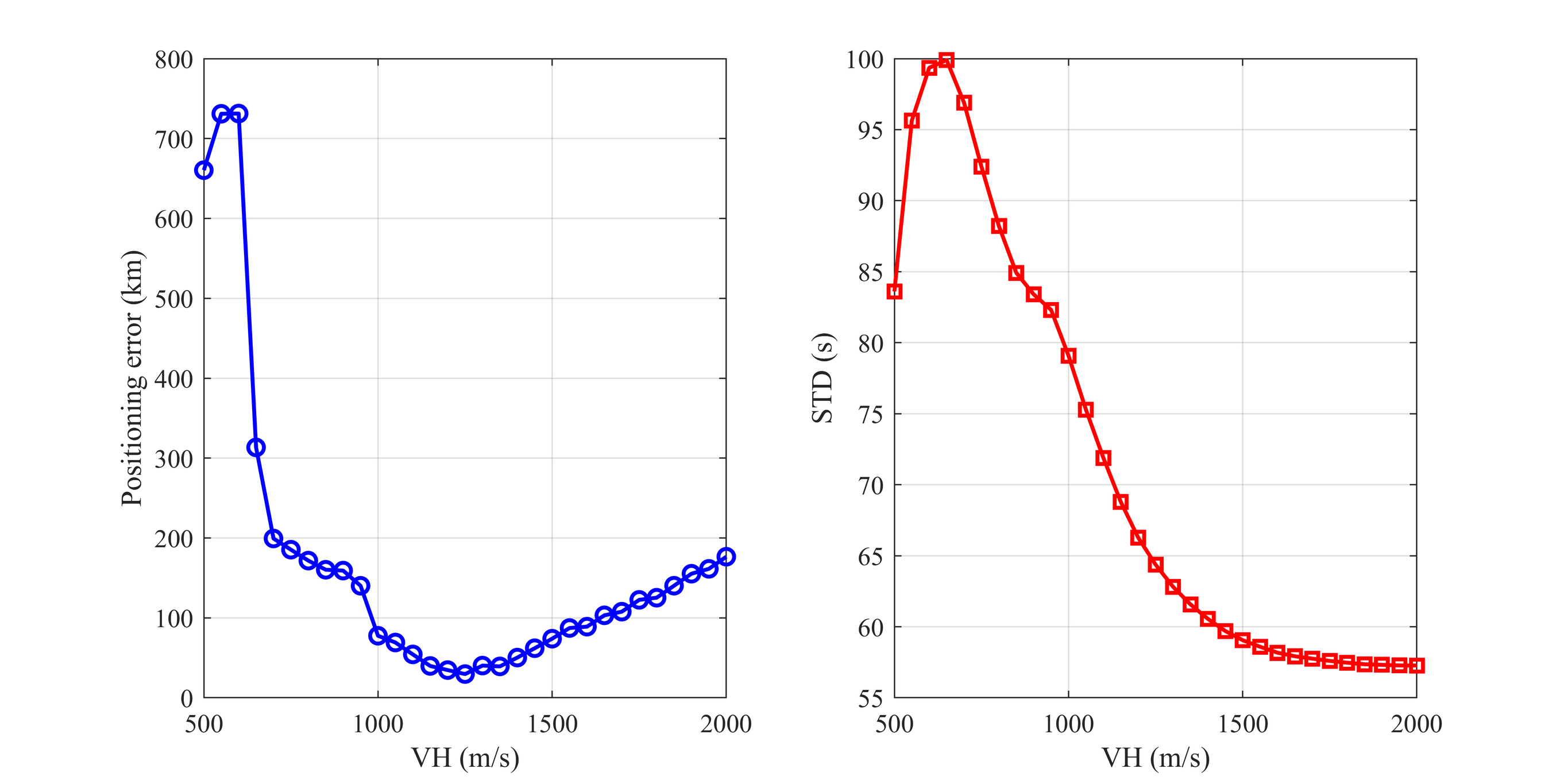}
\caption{Horizontal location error (left, in km) and weighted standard deviation of travel-time residuals STD (right, in seconds) as functions of the assumed horizontal propagation velocity $v_h$.}
\label{fig:error_vs_velocity}
\end{figure*}

\begin{figure*}
\centering
\includegraphics[width=0.8\textwidth]{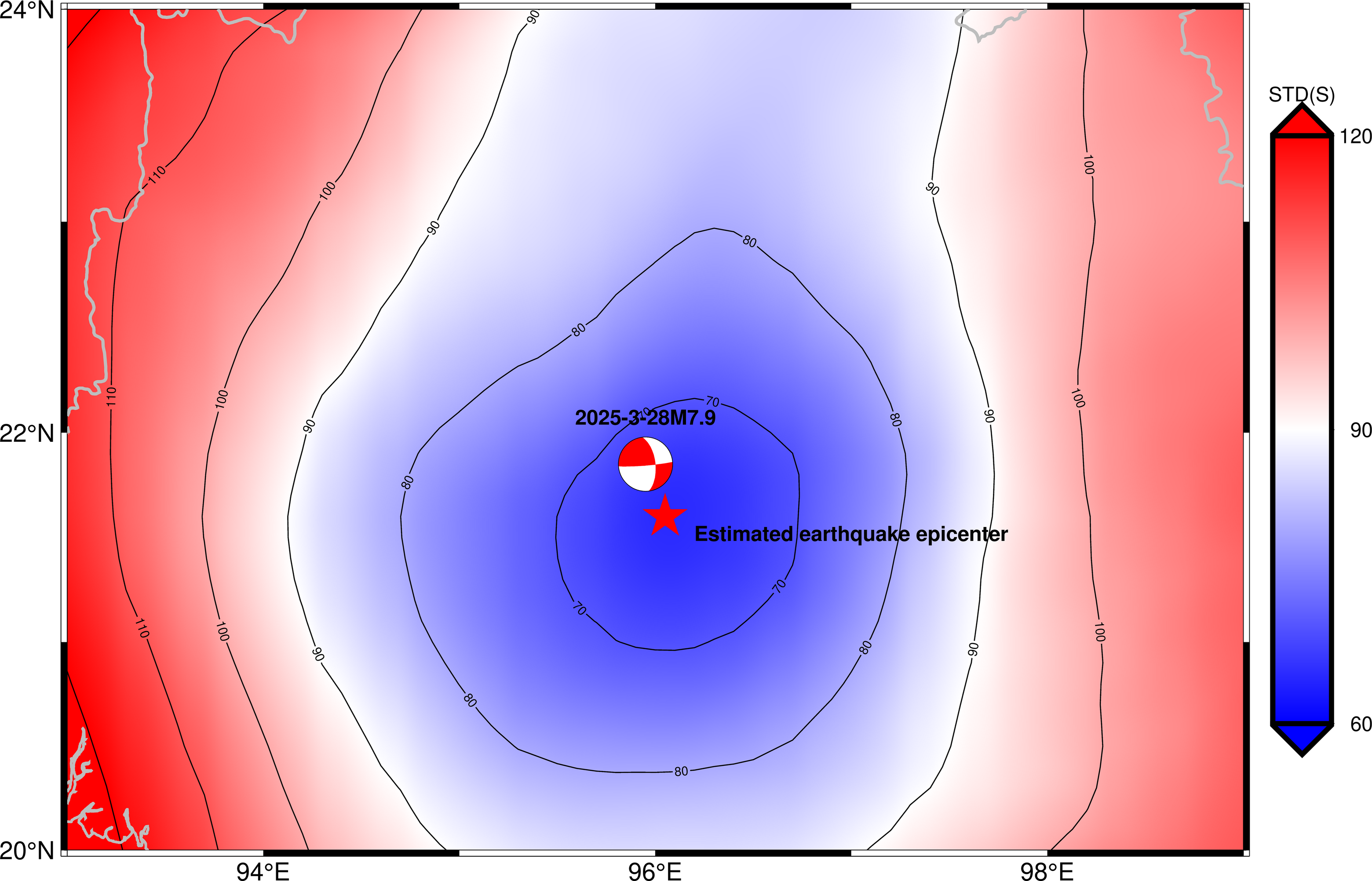}
\caption{Comparison of the estimated disturbance source location (blue star) with the actual earthquake epicenter (red star). The color shading shows the STD (s) of travel-time residuals for each candidate source location; the minimum-STD location represents the apparent centroid of the ionospheric disturbance energy.}
\label{fig:epicenter_location}
\end{figure*}

The error--velocity relationship curve from the grid search shows that the minimum localization error is achieved when the assumed horizontal propagation velocity is $v_h = 1210$ m s$^{-1}$ (Figure~\ref{fig:error_vs_velocity}). At this optimal velocity, the weighted standard deviation STD of the travel-time residuals is 21.5 s (Figure~\ref{fig:error_vs_velocity}), indicating a high degree of consistency among the observations. Note that STD is measured in seconds and quantifies the misfit in arrival time, not a spatial error. When $v_h$ deviates from the range of 1000--1500 m s$^{-1}$, the localization error increases sharply. The comparison between the inverted location and the actual epicenter is shown in Figure~\ref{fig:epicenter_location}. The optimal localization point (96.1°E, 21.6°N) shows a southeastward offset of 31.8 km relative to the true epicenter.

\subsection{Comparison of BeiDou GEO and Multi-GNSS Detections}
A direct comparison between the BeiDou GEO observations (Figure~\ref{fig:cid_azimuth_geo}) and the multi-GNSS results (Figure~\ref{fig:cid_azimuth_multisystem}) highlights the unique advantages of each. While multi-GNSS provides broader spatial coverage, revealing the full extent of the disturbance field, the BeiDou GEO data (e.g., Figure~\ref{fig:cid_bds_geo}c) offer superior temporal continuity at fixed points. This allows for a more precise determination of the wave arrival time and period at specific locations, which is challenging with the rapidly moving IPPs of MEO GNSS satellites. The consistency in the observed propagation velocity (1.2 km/s) and azimuthal patterns across both datasets validates our findings and demonstrates the synergistic power of combining these observation types.

\section{Discussion}
\subsection{Interpretation of Pre-Seismic LAIC Signatures}
The observed "positive-negative alternation-conjugate response-decay" sequence of the pre-seismic anomaly is phenomenologically consistent with the interplay of multiple plausible LAIC coupling paths rather than a single mechanism. One possible pathway for the initial positive anomaly at 06:00 UT is that microfractures associated with crustal stress accumulation generate infrasound disturbances that propagate upward and locally modify the ionospheric plasma recombination rate through collisional heating; however, this interpretation remains debated in the literature. The subsequent dipole-like anomaly distribution can be qualitatively explained by two concurrent mechanisms: (1) a quasi-static electric field possibly generated in the near-surface environment and mapped into the ionosphere, causing plasma redistribution via the E$\times$B drift; and (2) the compression and rarefaction phases of atmospheric gravity waves (AGWs) causing corresponding increases and decreases in F-layer electron density.

The cross-equatorial conjugate phenomenon observed at 10:00 UT is particularly suggestive. This structure may result from a combination of (1) plasma convection caused by field-aligned current transmission along geomagnetic field lines and (2) the propagation of AGWs released from the seismogenic region through the atmospheric waveguide to the conjugate region. The final decay phase, with an intensified negative anomaly, may be phenomenologically consistent with the late preparatory stage of the earthquake, but further statistical evidence from larger event samples would be required to establish any causal relationship. Although the observed sequence is suggestive of LAIC-related processes, it remains a single-case observation and should be regarded as suggestive rather than conclusive without further multi-event statistical verification.

\subsection{Governing Factors of CID Asymmetry}
The strong CID asymmetry, with energy focused to the southeast, can be explained by the joint control of three factors: the source mechanism, the propagation path, and the background environment.
\begin{itemize}
    \item \textbf{Source Mechanism}: The fault motion in this strike-slip earthquake was primarily horizontal shear. This motion generates more significant ground displacement and stress changes perpendicular to the fault, thereby exciting stronger AGWs in that direction.
    \item \textbf{Propagation Path}: The propagation of ionospheric disturbances is modulated by the geomagnetic field. At the epicenter, the magnetic declination was calculated to be 2.42° East. This geometry facilitates more efficient propagation of AGW energy along magnetic field lines in the southeast direction.
    \item \textbf{Background Environment}: The earthquake occurred near the equatorial anomaly region, where the daytime eastward electrojet (EEJ) drives plasma to drift eastward. The seismic disturbance superimposed on this background, where the electron density is already higher in the southeast direction, leads to more significant negative anomalies due to the rarefaction effects of AGWs. This combination results in the observed concentration of disturbance energy to the southeast.
\end{itemize}
The initial negative anomaly in the southeast may be caused by a rarefaction wave generated by instantaneous surface subsidence, leading to a decrease in ionospheric electron density. The smaller positive anomaly in the northwest could correspond to a compressional wave from surface uplift on the opposite side of the fault. The wave speed is determined by the medium properties, while the disturbance amplitude and polarity are dominated by the source mechanism and coupling path, resulting in consistent propagation speeds but varying anomaly characteristics \citep{Chen2024,Liu2015}.

\subsection{Physical Insights from Source Inversion}
The traditional grid search method can yield unrealistically large (thousand-kilometer) offsets when the IPP distribution is highly uneven. The spatial density-weighted inversion substantially mitigates this bias. For the Myanmar event, the minimum-STD location is offset by about 32 km to the southeast of the seismic epicenter. This offset should be interpreted as the apparent centroid of the ionospheric disturbance energy rather than the exact nucleation point of the earthquake. Importantly, the inversion no longer produces unrealistically large offsets even when the observational geometry is highly anisotropic.
Given the several-hundred-kilometre rupture length of this strike-slip earthquake and the spatio-temporal integration inherent in the ionospheric response, offsets of a few tens of kilometres between the seismic epicentre and the CID centroid are physically reasonable and compatible with previous observations of long-rupture events.

Crucially, the localization result shows a systematic southeastward offset relative to the actual epicenter. This offset is not an error but a physical signature, revealing the significant modulation effect of background ionospheric dynamics on the propagation of co-seismic disturbance energy. The eastward plasma drift driven by the equatorial electrojet effectively shifts the apparent source of the ionospheric disturbance to the southeast. Furthermore, the optimal horizontal propagation velocity of 1210 m/s derived from the inversion is significant. This speed is too fast for standard internal gravity waves but falls squarely within the typical range for acoustic--gravity waves propagating in the thermosphere, where the acoustic speed is elevated due to high temperatures (typically 800--1300 m/s). Our inferred propagation speed and dominant frequency band (2--8 mHz) are consistent with the characteristics of earthquake-generated acoustic--gravity waves revealed by recent modelling studies \citep[e.g.][]{Gao2023,Li2025,Zhang2025}, which further supports the AGW interpretation of the observed CIDs.

\subsection{Broader Context and Future Work}
This study outlines a phenomenologically self-consistent picture of the LAIC energy transfer chain for this strike-slip earthquake based on combined pre-seismic and co-seismic ionospheric observations. However, these findings are based on a single case study, and several limitations should be acknowledged. First, the GNSS receiving stations used here are distributed in a limited azimuthal range relative to the epicenter, so the inferred disturbance geometry is constrained primarily from the north and northeast. Although the spatial density-weighted inversion mitigates extreme biases under such anisotropic sampling, additional observations from other azimuths would be required to fully resolve the three-dimensional disturbance field. Second, the present study focuses on one large strike-slip event; future research should apply this methodology to a broader range of earthquake types and magnitudes to test the generalizability of the observed mechanisms. Future work should also focus on quantitative modeling to reproduce the observed CID amplitudes and asymmetry. This would require coupling a realistic seismic source model with a full-wave or analytical AGW model for the atmosphere \citep[e.g.][]{Gao2023,Zhang2025} and an ionospheric chemistry/dynamics model (like SAMI3), following the approach of recent earthquake-triggered AGW simulations \citep[e.g.][]{Li2025}. Such an effort could help to constrain key LAIC parameters, such as the efficiency of energy transfer from the ground to the ionosphere.

\section{Conclusions}
This study systematically analyzed the pre-seismic ionospheric anomalies and co-seismic disturbances (CIDs) from the 2025 M7.9 Myanmar strike-slip earthquake using high-resolution BeiDou GEO and multi-system GNSS data. We revealed key physical mechanisms of lithosphere-atmosphere-ionosphere (LAIC) coupling. The main conclusions are:

1. \textbf{Enhanced Detection Capability}: The fixed ionospheric pierce points of BeiDou GEO satellites, combined with multi-system GNSS, significantly improved the capability to reliably detect weak pre-seismic anomalies and rapid co-seismic evolution, mitigating the limitations associated with traditional MEO GNSS observations with rapidly moving IPPs.

2. \textbf{Pre-Seismic Anomaly Signature}: A significant negative TEC anomaly was identified three days before the earthquake. Its evolution—characterized by an initial positive anomaly, a dipole-like structure, a cross-equatorial conjugate response, and eventual decay—is consistent with multiple plausible LAIC coupling paths, including acoustic-gravity waves, quasi-static electric fields, and E$\times$B drift effects, but does not by itself constitute definitive proof of any single mechanism.

3. \textbf{Co-Seismic Disturbance Asymmetry}: CIDs were dominated by 2-8 mHz acoustic-gravity waves propagating at $\sim$1.2 km/s. A key finding is the significant asymmetric propagation, with disturbance energy concentrated in the southeast direction. This asymmetry is jointly controlled by the strike-slip fault mechanism (source), geomagnetic field orientation (path), and the daytime equatorial electrojet (background environment).

4. \textbf{Physically-Revealing Source Inversion}: An innovative spatial density-weighted inversion method substantially mitigates the bias caused by the highly uneven IPP distribution. For the Myanmar event, the minimum-STD location is offset by about 32 km to the southeast of the seismic epicenter. Given the large rupture length of this strike-slip earthquake, this offset should be interpreted as the apparent centroid of the ionospheric disturbance energy rather than the exact nucleation point. Importantly, the inversion no longer yields unrealistically large (thousand-kilometer) offsets even under highly anisotropic IPP geometries.

This study outlines a phenomenologically self-consistent picture of the LAIC energy transfer chain for this strike-slip earthquake based on combined pre-seismic and co-seismic ionospheric observations. The findings deepen the physical understanding of seismo-ionospheric coupling and provide new methods for space-based earthquake monitoring and rapid response.

\begin{acknowledgments}
This work was supported in part by the Natural Science Foundation of China (NSFC), grant number 42274108, the Central Public-interest Scientific Institution Basal Research Fund (No. CEAIEF2025030105) and Natural Science Basic Research Program of Shaanxi Province (Program No.2025JC-YBQN-448). The data processing process used the IonKit-NH software (\url{https://doi.org/10.1016/j.eqrea.2024.100353}).
\end{acknowledgments}

\begin{dataavailability}
The GNSS-TEC data used in this study are available at Figshare via \url{https://doi.org/10.6084/m9.figshare.28847789.v1}.
\end{dataavailability}


\bibliographystyle{gji}

\clearpage

\clearpage

\clearpage

\clearpage

\clearpage

\clearpage

\clearpage

\clearpage

\clearpage

\clearpage

\label{lastpage}

\end{document}